\documentclass[letter,traditionalabstract,onecolumn]{aa} 

\usepackage{graphicx}
\usepackage{txfonts}
\usepackage[version=4]{mhchem} 
\usepackage{xspace}
\newcommand{\HC}{H$_2$CCCH$^+$\xspace}
\newcommand{\PI}{C$_3$H$_2$\xspace}

\begin{document}

   \title{
    Discovery of \HC in TMC-1 \thanks{Based on observations carried out
with the Yebes 40m telescope (projects 19A003,
20A014, 20D15, and 21A011) and the Institut de Radioastronomie Millim\'etrique (IRAM) 30m telescope. The 40m
radiotelescope at Yebes Observatory is operated by the Spanish Geographic Institute
(IGN, Ministerio de Transportes, Movilidad y Agenda Urbana). IRAM is supported by INSU/CNRS
(France), MPG (Germany) and IGN (Spain).}}

   \subtitle{}

   \author{W. G. D. P. Silva \inst{1}  \and
          J.~Cernicharo  \inst{2}  \and
          S.~Schlemmer \inst{1} \and
          N.~Marcelino \inst{3,4} \and
          J.-C.~Loison \inst{5} \and
          M.~Ag\'undez  \inst{2}  \and
          D.~Gupta  \inst{1}  \and
          V.~Wakelam \inst{6} \and
          S.~Thorwirth  \inst{1}   \and
          C.~Cabezas \inst{2} \and
          B.~Tercero  \inst{3,4} \and
          J.~L.~Dom\'enech  \inst{7} \and
          R.~Fuentetaja \inst{2}  \and
          W.-J.~Kim  \inst{1}  \and
          P.~de~Vicente \inst{3}  \and
          O.~Asvany  \inst{1}  
          }

   \institute{I. Physikalisches Institut, Universit\"at zu K\"oln, Z\"ulpicher Str. 77, 50937  K\"oln, Germany\\
              \email{asvany@ph1.uni-koeln.de}
         \and
 Dept. de Astrofísica Molecular, Instituto de Física Fundamental (IFF-CSIC), Serrano 121, 28006 Madrid, Spain
             \and
Centro de Desarrollos Tecnológicos, Observatorio de Yebes (IGN), 19141 Yebes, Guadalajara, Spain
             \and
Observatorio Astronómico Nacional (OAN, IGN), Madrid, Spain
 \and
Institut des Sciences Mol\'eculaires (ISM), CNRS, Univ. Bordeaux, 351 cours de la Lib\'eration, 33400, Talence, France
\and
Laboratoire d'Astrophysique de Bordeaux, Univ. Bordeaux, CNRS, B18N, all\'ee Geoffroy Saint-Hilaire, 33615 Pessac, France
\and
Instituto de Estructura de la Materia, IEM-CSIC, Serrano 123, 28006 Madrid, Spain   
}

   \date{Received X, 2023; accepted X, 2023}

 
  \abstract
   {Based on a novel laboratory method, 14  mm-wave lines of the molecular
   ion \HC  have been measured in high resolution, and the spectroscopic constants 
   of this asymmetric rotor determined with high accuracy.
   Using the Yebes 40 m and IRAM 30 m radio telescopes, we detect  
   four lines  of \HC towards the cold dense core TMC-1. 
   With a dipole moment of about 0.55~Debye obtained from high-level ab initio calculations, we derive 
   a column density of 5.4$\pm$1$\times$10$^{11}$ cm$^{-2}$ and 1.6$\pm$0.5$\times$10$^{11}$ cm$^{-2}$ for 
   the ortho and para species, respectively, and an abundance ratio N(H$_2$CCC)/N(\HC)= 2.8$\pm$0.7.
The chemistry of \HC is modelled using the most recent chemical
network for the reactions involving the formation of \HC. We find a reasonable agreement 
between model predictions and observations, and new insights into the chemistry
of C$_3$ bearing species in TMC-1 are obtained.}

   {}

   \keywords{astrochemistry --
                ISM: molecules --
                ISM: individual objects: TMC-1 --
                line: identification --
                molecular data --
                methods: laboratory: molecular
               }

   \maketitle
%

\section{Introduction}

Molecular ions are important intermediates in the chemistry of the interstellar medium (ISM). 
These charged species can rapidly react with neutral partners or recombine with electrons to form other 
ionic and neutral molecules under astrophysical conditions \citep{lar12,agu2013}.
Despite their fundamental role in astrochemistry, many ions remain elusive mainly due to their 
highly reactive character and lack of accurate laboratory data to support 
astronomical detections \citep{mcg20}. Consequently, many ions in astrochemical models 
and theories await confirmation through spectroscopic detection in the ISM.

One example is the formation of the hydrocarbons \PI and C$_3$H, whose both 
cyclic (c-\PI and c-C$_3$H) and linear (l-\PI and l-C$_3$H, i.e.  H$_2$CCC and HCCC, respectively) isomers were 
detected in the ISM \citep{Thaddeus1985_c3h,Thaddeus1985_c3h2,Yamamoto1987,Cernicharo1991}, 
and even deuterated versions were observed \citep{Bell1986,Spezzano2013,Spezzano2016,Agundez2019}.
The synthesis of the cyclic and linear forms of \PI and C$_3$H is thought 
to occur via the dissociative recombination of the respective 
isomers of C$_3$H$_3$$^+$ with electrons, c-C$_3$H$_3$$^+$ and \HC \citep{Maluendes1993}. In turn, both isomers 
of C$_3$H$_3$$^+$ would be produced through the radiative association of C$_3$H$^+$ and H$_2$ \citep{sav05}. The proof of this chemical pathway 
for the cyclic variants is difficult because c-C$_3$H$_3$$^+$ is a symmetric molecule 
and can only be detected based on its rovibrational fingerprints in the infrared \citep{zha14}, which might be feasible with the James Webb Space Telescope (JWST) in the near future. While the singly-deuterated version c-C$_3$H$_2$D$^+$ could be probed by radio astronomy, it has a predicted low dipole moment and low column densities \citep{gupta23}. This  leaves 
only \HC as a good candidate for radio astronomical searches.



In this Letter, based on a novel experimental method, 
we report first laboratory  mm-wave data of \HC and its 
radio-astronomical detection towards the cold dark core TMC-1.
We derive its column density towards TMC-1
and discuss these results in the
context of state-of-the-art chemical models.

\section{Laboratory Work}

\begin{figure}
 \includegraphics[width=0.48\textwidth]{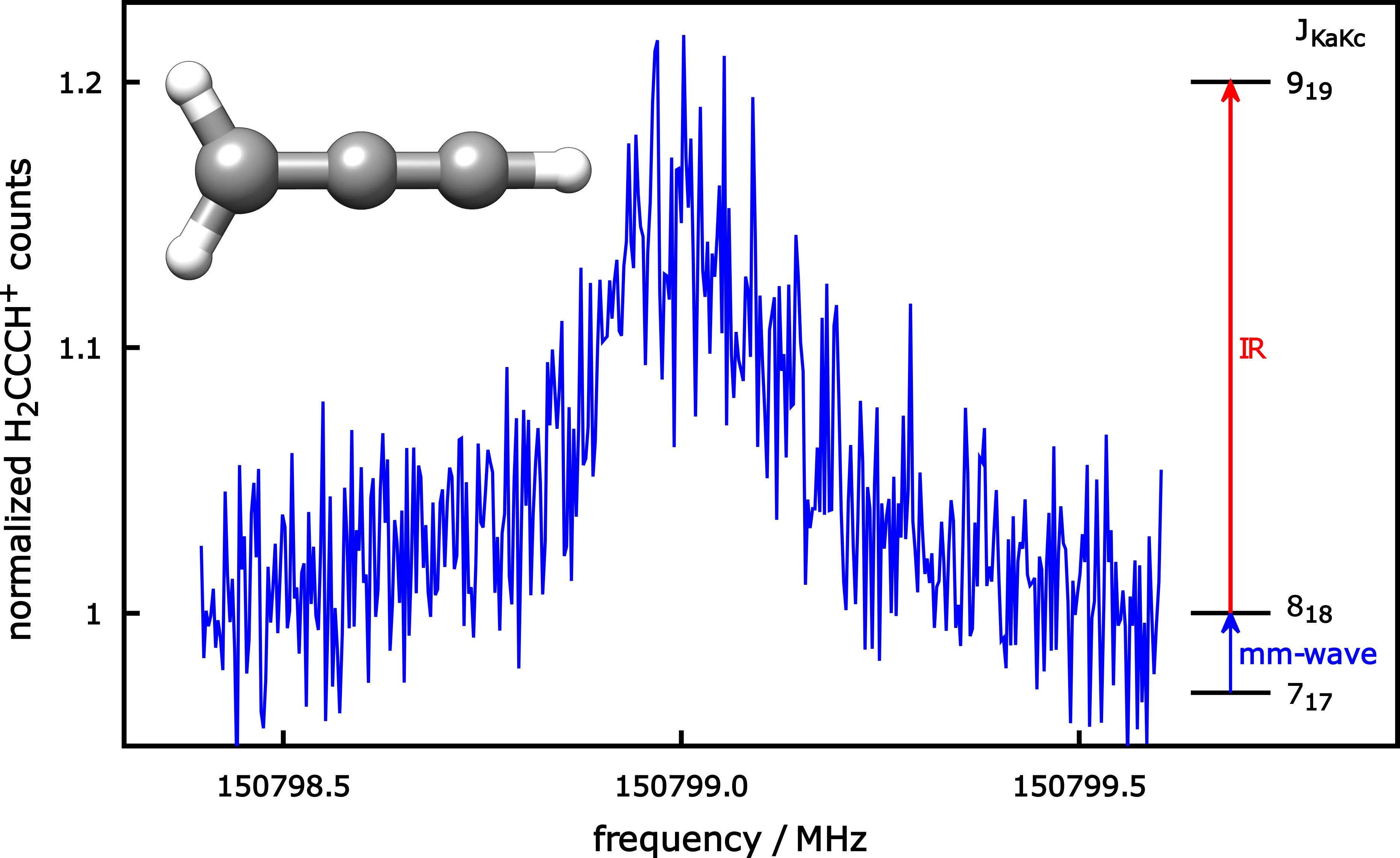}
 \caption{\label{fig:rot} Pure rotational transition ($J'_{Ka'Kc'}$ $\leftarrow$ $J''_{Ka''Kc''}$ = $8_{18}$ $\leftarrow$ $7_{17}$) 
 of \HC recorded using the double resonance spectroscopic scheme. 
 For this measurement, the IR laser frequency (red arrow)
 was kept fixed on resonance with the  $9_{19}$ $\leftarrow$ $8_{18}$ rovibrational transition within the $\nu_{1}$ band.
 }
\end{figure}

\HC is a closed-shell,  planar and near-prolate asymmetric top molecular ion 
(see sketch in Fig. 1). \HC ions were generated in the Cologne laboratory 
in a storage ion source via electron impact ionization ($E_e$  $\approx$30 eV) of 
the precursor gas allene (C$_3$H$_4$). By applying a novel trap-based technique 
called leak-out-spectroscopy (LOS, \citealp{scm22a})
in the cryogenic ion trap machine COLTRAP \citep{asv10,asv14}, 
the  vibrational bands
$\nu_1$ and $\nu_3$+$\nu_5$ were measured in the range 3180 - 3240~cm$^{-1}$ in high resolution. 
The vibrational measurements, whose details will be described in a forthcoming publication, 
enabled the ground state spectroscopic parameters of \HC to be determined.
Subsequently, pure rotational lines were detected using a vibrational-rotational double resonance (DR) method.
Such methods have  been reviewed by \cite{asv21d},
and the particular scheme involving LOS has only recently been demonstrated by \cite{asv23}.
An example measurement for \HC is shown in Fig.~\ref{fig:rot}.

DR spectra were recorded in multiple individual measurements in which 
the mm-wave frequency (blue arrow in Fig.~\ref{fig:rot}) 
was stepped in an up-and-down manner several times. 
Selected rovibrational lines from the
$\nu_1$ or the $\nu_3+\nu_5$ combination band 
 were used for the IR excitation (red arrow in Fig.~\ref{fig:rot}).
The frequency steps of the mm-wave radiation were 
kept constant in individual experiments, and varied between 3 and 50~kHz
(the larger steps typically used for searching new lines).
The spectroscopic data were normalized 
employing a frequency switching procedure, i.e., by
dividing the \HC counts monitored while scanning the spectral range of interest
by the counts at an off-resonant mm-wave reference frequency.
Therefore, the baseline in  Fig.~\ref{fig:rot} is close to unity.  
The on-resonance signal enhancement  is on the order of 15~\%. 
Transition frequencies were determined by adjusting the parameters 
of an appropriate line shape function  (typically a Gaussian) to 
the experimental spectrum in a least-squares procedure.
In total, 14 rotational lines were detected in the laboratory and are summarised in Table~\ref{rotlines}.
The frequencies and their uncertainties in the table
result from the weighted average of several (up to eleven) 
independent line-center determinations for each transition.

\begin{table}
\small
\begin{center}
\caption{\label{rotlines} Ground state rotational transition frequencies  of \HC,
 determined by laboratory and astronomical detections. 
 The uncertainty in the last digits of each line is provided in parentheses.} 
\begin{tabular}{c@{ $\leftarrow$ }c r@{}l r@{}l l}
\hline
$J'_{Ka'Kc'}$ & $J''_{Ka''Kc''}$  & \multicolumn{2}{c}{frequency / MHz}   & \multicolumn{2}{c}{obs-calc / kHz} & source \\
\hline
$  2_{12} $ & $ 1_{11} $   &    37702&.627(10)  &  0 &.3 & astro $^a$\\
$  2_{02} $ & $ 1_{01} $   &    38037&.044(10)  &  7 &.0 & astro $^a$\\  
$  2_{11} $ & $ 1_{10} $   &    38368&.581(20)  & 33 &.0 & astro $^a$\\ 
$  5_{15} $ & $ 4_{14} $   &   94254&.055(4)    &   7&.7  & lab  \\
$  5_{05} $ & $ 4_{04} $   &   95086&.000(10)   &  $-$12&.8 & lab    \\
$  5_{14} $ & $ 4_{13} $   &   95918&.753(3)    &    7&.9  & lab   \\
$  6_{16} $ & $ 5_{15} $   &   113103&.272(4)   &   $-$2&.6  & lab   \\
$  6_{06} $ & $ 5_{05} $   &   114099&.086(6)   &     6&.0 & lab  \\
$  6_{15} $ & $ 5_{14} $   &   115100&.840(3)   &   $-$2&.7   & lab  \\
$  7_{17} $ & $ 6_{16} $   &   131951&.635(3)   &   $-$5&.4  & lab   \\
$  7_{07} $ & $ 6_{06} $   &   133109&.888(6)   &  $-$3&.9 & lab  \\
$  7_{16} $ & $ 6_{15} $   &   134282&.033(4)   &  $-$5&.1 & lab    \\
$  8_{18} $ & $ 7_{17} $   &   150799&.008(4)   &    5&.3   & lab  \\
$  8_{08} $ & $ 7_{07} $   &   152118&.074(5)   &    0&.8 & lab  \\
$  8_{17} $ & $ 7_{16} $   &   153462&.176(3)   &    $-$3&.7   & lab  \\
$ 10_{1 10} $ & $ 9_{19} $  &   188490&.152(12) &    1&.6  & lab   \\
$ 10_{19} $ & $ 9_{18} $   &   191818&.719(10)  &    27&.3  & lab   \\
\hline                                                  
\end{tabular}                                      
\end{center}       
\tablefoot{\\
\tablefoottext{a}{Observed frequency assuming a Local Standard of Rest
velocity (v$_{LSR}$) of 5.83 km\,s$^{-1}$ \citep{Cernicharo2020a}.}}

\end{table}  

The fit of the assigned lines (lab + astro) was carried out using  Watson's $S$-reduced 
Hamiltonian in the \emph{I$^r$} representation as implemented in Western's PGOPHER program \citep{wes17}. 
The resulting spectroscopic parameters are given in Table~\ref{tab:fit}.
As \HC is a near-prolate asymmetric top ($\kappa=-0.9976$) with an $a$-type spectrum, the $A_0$ rotational constant 
is not well constrained from experiment.
Overall, the experimental rotational and centrifugal distortion constants
compare favorably with those calculated by \cite{hua11a} and very well with the
best estimate values obtained in this study (given in the last three columns of Table~\ref{tab:fit}).
The obtained obs-calc values for the rotational lines are given in Table~\ref{rotlines}. In total, 
the  weighted rms of the fit is on the order of 1.4, indicating somewhat optimistic 
uncertainties of our measurements.

\begin{table*}[h]
\small
\centering
\caption{\label{tab:fit} Spectroscopic parameters of \HC (in MHz)
obtained by fitting Watson's S-reduced Hamiltonian to the rotational transitions from Table~\ref{rotlines}.
Western's PGOPHER program \citep{wes17} has been used for the fit. 
The experimental uncertainties are reported in parentheses.
}
\begin{tabular}{lr@{}lr@{}lr@{}lr@{}l}
\hline
                     &  \multicolumn{2}{c}{Exp.}               &     \multicolumn{6}{c}{Calculated}                       \\   
                     \cline{4-9}
Parameter            &  \multicolumn{2}{c}{This study} &  \multicolumn{2}{c}{\cite{hua11a} }   & \multicolumn{2}{c}{This study$^a$} & \multicolumn{2}{c}{This study, BE$^b$}   \\ 
\hline
  $A_0$                &   281856 & .(247)      &   281911 & .9                              &  283318 & .43                  &  282182 & .6                                      \\ 
  $B_0$                &     9675 & .841(1)    &    9580  &.2                               &    9670 & .18                  &    9675 & .97                                     \\
  $C_0$                &     9342 & .877(1)    &    9251  &.9                               &    9337 & .82                  &    9343 & .14                                     \\
  $D_J\times 10^3$   &        2 & .942(4)      &       3  &  .                              &       2 & .68                  &       3 & .06                                      \\ 
  $D_{JK}$           &        0 & .4388(9)     &       0  &.479                             &       0 & .479                 &       0 & .433                                   \\ 
  $D_K$              &       20 & .678$^c$     &      20  &.862                             &      20 & .678                 &      20 & .678                                 \\ 
  $d_1\times 10^3$   &     $-$0 & .121(3)      &       0  &   .                              &       $-$0 & .097             &    $-$0 & .103                                   \\   
  $d_2\times 10^3$   &     $-$0 & .051$^c$     &       0   &  .                               &       $-$0 & .037            &    $-$0 & .051                                            \\ 
  $\mu_A$ / Debye    &     \multicolumn{2}{c}{$\cdots$}             &         0&.524  $^d$                       &         0 & .55                      &       \multicolumn{2}{c}{$\cdots$}                                           \\
\hline
\end{tabular}
\tablefoot{\\
\tablefoottext{a}{From equilibrium structure and force field calculated at the ae-CCSD(T)/cc-pwCVQZ and fc-CCSD(T)/ANO1 levels, respectively.}\\
\tablefoottext{b}{Best estimate (BE) values, obtained through scaling of calculated values via isoelectronic propadienylidene, \ce{H2CCC}, see text for details.}\\
\tablefoottext{c}{Fixed to best estimate value.}
\tablefoottext{d}{From \cite{hua11}}\\
}
\end{table*}


   \begin{figure}
   \centering
\includegraphics[width=0.45\textwidth]{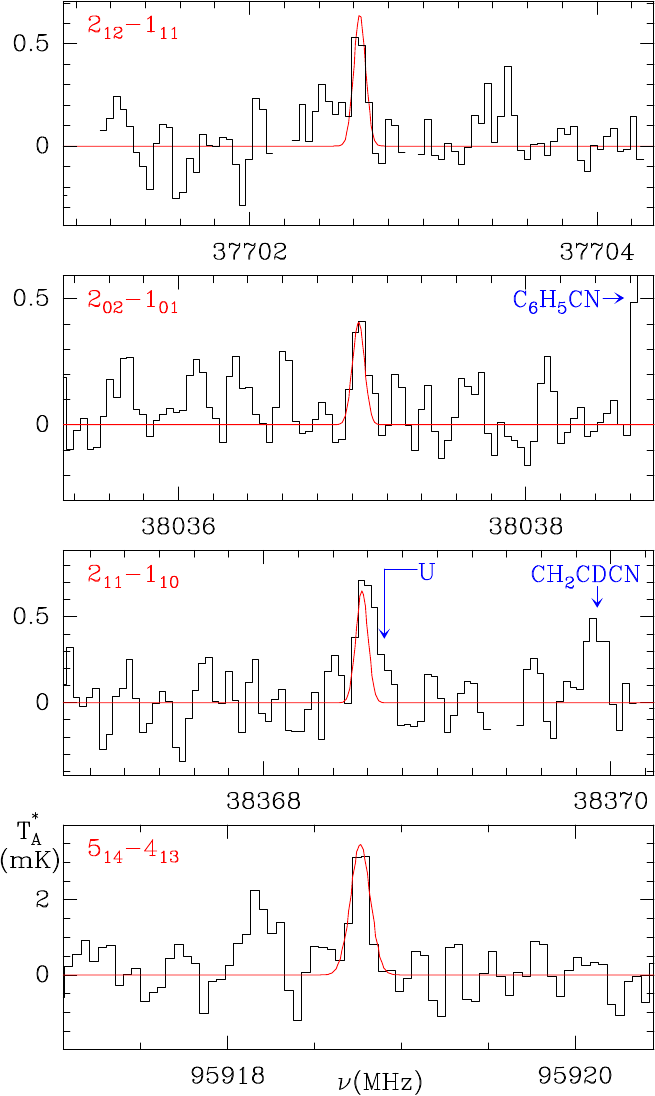}
   \caption{Lines of \HC observed in the 31–115~GHz
domain towards TMC-1. The abscissa corresponds to the rest frequency in MHz.
 The ordinate is the antenna temperature corrected for atmospheric and telescope losses in mK. The spectral resolution is 38~kHz below 50 GHz and 48 kHz above. Line parameters are given in Table \ref{Table:Line_Parameters}. The red lines show the computed synthetic spectra (see text).}
\label{FigGam}%
    \end{figure}

\section{Quantum Chemical Calculations}

The \HC molecular ion has been a subject of several quantum-chemical investigations
in the past \citep[e.g.,][and references therein]{Botschwina_JChemSocFaradayTrans_89_2219_1993,botschwina_PCCP_13_7921_2011,hua11a,marimuthu_JMS_374_111377_2020}. In the present study, complementary high-level calculations were
performed at the CCSD(T) level of theory \citep{raghavachari_chemphyslett_157_479_1989}
together with correlation consistent (augmented) polarized weighted core-valence basis sets \citep{kendall_JCP_96_6796_1992,peterson_JCP_117_10548_2002} as well as atomic natural orbital basis sets
\citep{almlof_JCP_86_4070_1987}.
All calculations were performed 
using the CFOUR program suite \citep{cfour_JCP_2020,harding_JChemTheoryComput_4_64_2008}.
Equilibrium rotational constants were calculated at the
all-electron \mbox{(ae-)CCSD(T)/cc-pwCVQZ} level of theory
that is known to yield molecular equilibrium structural parameters of very high quality for molecules comprising first- and second-row elements \citep[e.g.,][]{coriani_JCP_123_184107_2005}. 

Zero-point vibrational contributions 
$\frac{1}{2} \sum_{i} \alpha_i^{A,B,C,calc}$ ($=\Delta A_0, \Delta B_0, \Delta C_0$)   
to the equilibrium rotational constants and centrifugal distortion parameters were
calculated at the frozen core (fc-)CCSD(T)/ANO1 level. 
Best estimate (BE, Tables~\ref{tab:fit} and \ref{tab:a1}) rotational and centrifugal distortion constants were finally obtained through
empirical scaling of the calculated rotational parameters 
using factors (i.e., the ratios $X_{exp}/X_{calc}$ of
a given parameter)
derived from isoelectronic propadienylidene, H$_2$CCC, the pure rotational spectrum
of which is known well from previous study \citep{vrtilek_ApJL_364_L53_1990}. 
The technique of empirical scaling using structurally closely related 
(isoelectronic) species known from experiment may provide rotational parameters at a predictive power greatly
exceeding that of high-level calculations alone \citep[see, e.g.,][]{thorwirth_JCP_184308_2005,thorwirth_MolPhys_118_e1776409_2020,martinez_JCP_138_094316_2013},
and has been used recently to identify species
not studied in the laboratory using radio astronomy 
(see, e.g., \citealp{Cernicharo2020b}).

The dipole moment of \HC is not very large. At the \mbox{ae-CCSD(T)/aug-cc-pwCVQZ} level of theory the 
(center-of-mass frame)
equilibrium value $\mu_e$ amounts to 0.53\,D in good agreement with earlier estimates.
Zero-point vibrational effects (fc-CCSD(T)/ANO1) have an almost negligible influence resulting in $\mu_0=0.55$\,D.


\section{Observations}
New receivers, built within the Nanocosmos\footnote{ERC grant ERC-2013-Syg-610256-NANOCOSMOS.\\
https://nanocosmos.iff.csic.es/} project and installed at the Yebes 40 m radio telescope, were used
for the observations of TMC-1 ($\alpha_{J2000}=4^{\rm h} 41^{\rm  m} 41.9^{\rm s}$ and $\delta_{J2000}=
+25^\circ 41' 27.0''$).  
The observations of TMC-1 belong to the on-going QUIJOTE\footnote{\textbf{Q}-band \textbf{U}ltrasensitive \textbf{I}nspection \textbf{J}ourney to the \textbf{O}bscure \textbf{T}MC-1 \textbf{E}nvironment} 
line survey \citep{Cernicharo2021,Cernicharo2023}.
A detailed description of the telescope, receivers, and backends is 
given by \citet{Tercero2021}. Briefly, the receiver consists of two cold high electron 
mobility transistor amplifiers covering the 31.0-50.3 GHz band with horizontal and vertical             
polarizations. The backends are $2\times8\times2.5$ GHz fast Fourier transform spectrometers
with a spectral resolution of 38.15 kHz providing the coverage of the whole Q-band in both polarisations.

The observations, carried out during different observing runs, are 
 performed using the frequency-switching mode with a frequency throw of 10 MHz in the 
 very first observing runs, during November 2019 and February 2020, 8 MHz during the 
 observations of January-November 2021, and alternating these frequency throws 
 in the last observing runs between October 2021 and February 2023. 
 The total on-source telescope time is 850 hours
 in each polarization (385 and 465 hours for the 8 MHz and 10 MHz frequency throws, respectively).
 The sensitivity of the QUIJOTE line survey varies between 0.17 and 0.25\,mK in 
the 31-50.3\,GHz domain. The intensity scale used in this work, antenna temperature
($T_A^*$), was calibrated using two absorbers at different temperatures and the
atmospheric transmission model ATM \citep{Cernicharo1985, Pardo2001}.
Calibration uncertainties have been adopted to be 10~\%.
The beam efficiency of the Yebes 40 m telescope in the Q-band is given as
a function of frequency by $B_{\rm eff}$= 0.797 exp[$-$($\nu$(GHz)/71.1)$^2$]. The
forward telescope efficiency is 0.97. The telescope beam size varies from 56.7$''$ at 31 GHz to 35.6$''$ at 49.5 GHz.

The data of TMC-1 taken with the IRAM 30 m telescope consist of a 3 mm line survey obtained 
with the old ABCD receivers connected to an autocorrelator that provided a spectral resolution of 40 kHz \citep{Marcelino2007, Cernicharo2012a}. Some additional high sensitivity frequency windows observed in 2021 used the new 3\,mm EMIR dual polarization receiver connected to four fast Fourier transform spectrometers 
providing a spectral resolution of 49 kHz \citep{Agundez2022,Cabezas2022}. 
All the observations were performed using the frequency switching method.
The final 3\,mm line survey
has a sensitivity of 2-10 mK. However, at some selected frequencies the sensitivity is as low as 0.6 mK.


\section{Detection of H$_2$CCCH$^+$ in TMC-1}
We searched for one para (2$_{02}$-1$_{01}$) and two ortho (2$_{12}$-1$_{11}$, 2$_{11}$-1$_{10}$) lines of H$_2$CCCH$^+$ within the
QUIJOTE line survey. The three lines are clearly detected and
are shown in Fig. \ref{FigGam}. In the data at 3 mm we covered the
frequencies of three para and four ortho lines, with upper levels $J$= 4, 5, 6 and energies below 30 K. Only one line, the $J=5_{14}-4_{13}$, falls in one of the high sensitivity
windows  ($\sigma$= 0.6 mK) of our line survey, and it is also clearly detected (see Fig. \ref{FigGam}).
Two other lines are within frequency ranges with $\sigma$ below 2 mK, and are marginally detected.
The derived line parameters for all searched transitions of H$_2$CCCH$^+$ are given in Table \ref{Table:Line_Parameters}. We checked that the detected lines cannot be assigned to lines
of other species or isotopologues by exploring the spectral catalogues MADEX \citep{Cernicharo2012b},
CDMS \citep{Muller2005} and JPL \citep{Pickett1998}.

To estimate the column density of H$_2$CCCH$^+$, we considered the ortho and para levels as belonging to two different species. 
For the dipole moment we use the value of 0.55~D calculated in this work.
No collisional rates are
available for this molecule. However, \cite{Khalifa2019} have computed the collisional rates
between He and H$_2$CCC. This cumulenic species is isoelectronic with our molecule and
has a very similar structure. We therefore adopted these rates, correcting for the abundance of He with respect to H$_2$, to estimate the excitation temperatures of the  observed transitions of H$_2$CCCH$^+$. Assuming a volume density of (1-3)$\times$10$^4$ cm$^{-3}$ 
\citep{Fosse2001,Lique2006,Pratap1997}, we derive excitation temperatures close to 10 K for the 
$J$= 2-1 lines, and $\sim$8-10 K for the lines in the 3 mm domain, the largest value corresponding to
$n$(H$_2$)= 3$\times$10$^4$ cm$^{-3}$. These excitation temperatures are considerably 
larger than those obtained for H$_2$CCC, which are between 4 and 5 K, due to the larger dipole moment of this species
(4.1 versus 0.55\,D). From the adopted rotational temperatures of 9\,K, and assuming a
source of uniform brightness temperature with a radius of 40$''$  \citep{Fosse2001},
we derive a column density for
ortho-H$_2$CCCH$^+$ of (5.4$\pm$1)$\times$10$^{11}$ cm$^{-2}$. For the para species the estimated
column density is (1.6$\pm$0.5)$\times$10$^{11}$ cm$^{-2}$, a value that is consistent with the expected
ortho to para ratio of 3/1. The computed synthetic spectra show an
excellent agreement with the observed line intensities (see Fig.~\ref{FigGam}). Hence, the total column density of \HC in TMC-1 is (7.0$\pm$1.5)$\times$10$^{11}$ cm$^{-2}$.

It is interesting to compare the abundance of H$_2$CCCH$^+$ to that of H$_2$CCC. For the latter species we detected, with an excellent signal to noise ratio, all its ortho and para lines in the frequency range of our line surveys. The derived line parameters are summarized in Table \ref{Table:Line_Parameters_H2C3}
and the lines are shown in Fig.~\ref{FigH2C3}. The decline of the line intensity between the $J_u$= 2 
and $J_u$= 5 lines is obvious in Fig.~\ref{FigH2C3}, which indicates that the lines are not thermalized to the
kinetic temperature of the cloud for this species.
Using the collisional rates of \cite{Khalifa2019}, and
adopting the same assumptions on the source size than for H$_2$CCCH$^+$, we
derive a column density for the ortho and para species of 
(1.5$\pm$0.1)$\times$10$^{12}$ and (0.45$\pm$0.05)$\times$10$^{12}$ cm$^{-2}$, respectively.
The best fit is obtained for a density  of $n$(H$_2$)=8$\times$10$^3$ cm$^{-3}$. The total column density of
H$_2$CCC is (1.95$\pm$0.15)$\times$10$^{12}$ cm$^{-2}$ and 
the ortho to para ratio for this species is 3.3$\pm$0.6. The H$_2$CCC/H$_2$CCCH$^+$ abundance
ratio is 2.8$\pm$0.7 which is on the order of that found for C$_3$O/HC$_3$O$^+$ \citep{Cernicharo2020b}, but much smaller than the abundance ratio found in TMC-1 for other neutral 
species and their protonated forms \citep{Marcelino2020, Cernicharo2021b, Cernicharo2021c, Cabezas2022b, Agundez2022}.

   \begin{figure}
   \centering
\includegraphics[width=0.45\textwidth]{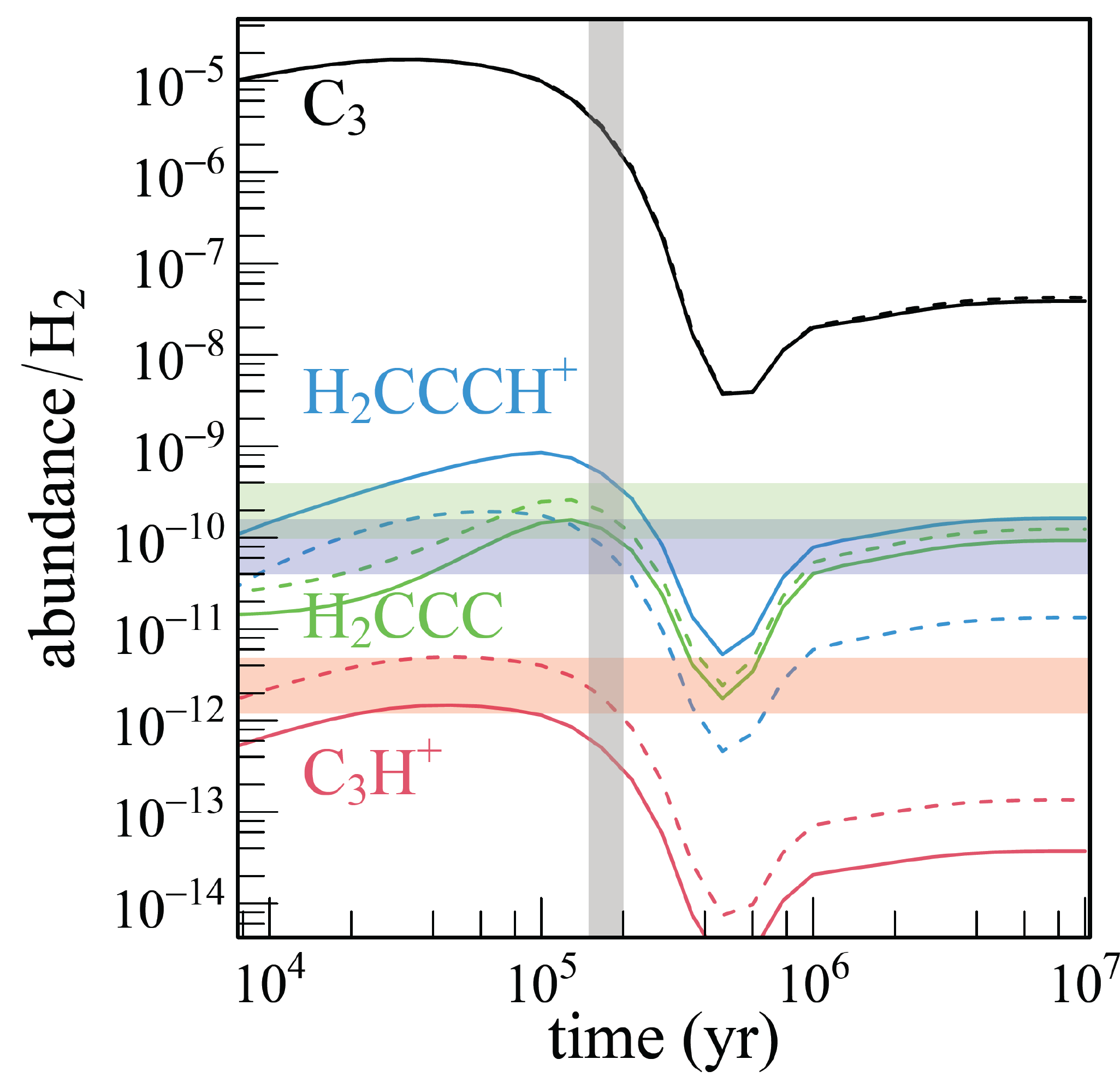}
   \caption{Continuous lines: Abundances of C$_3$, C$_3$H$^+$, H$_2$CCC and H$_2$CCCH$^+$ as a function of time predicted by our model. The horizontal colored areas represent the observations in TMC-1 assuming an uncertainty of 3 on the measurements (\citealt{RN11344} for C$_3$H$^+$ and this work for H$_2$CCCH$^+$). Dashed lines: the same with a threefold decrease in H$_2$ + C$_3$H$^+$ and a fivefold increase in H$_2$CCCH$^+$ + e$^-$ rate coefficients. The vertical grey area represents the values given by the most probable chemical age for TMC-1 given by the better agreement between calculations and observations for 67 key species.}
\label{Fig:model}%
    \end{figure}


\section{Discussion}

To describe the chemistry of \HC, we used the Nautilus code \citep{RN7545}, a 3-phase (gas, dust grain ice surface, and dust grain ice mantle) time-dependent chemical model with a chemical network for C$_3$H$_x^+$ species very similar to the one presented in \cite{RN8363}. To describe the physical conditions in TMC-1, we use an homogeneous cloud with a density equal to 2.5$\times$10$^4$ cm$^{-3}$, a temperature equal to 10~K for both the gas and the dust, a visual extinction of 30 mag and a cosmic-ray ionization rate of 1.3$\times$10$^{-17}$ s$^{-1}$. All elements are assumed to be initially in atomic form, except for hydrogen, which is entirely molecular \citep{RN3230}. The calculated abundances relative to H$_2$ for H$_2$CCC and H$_2$CCCH$^+$, and also for C$_3$ and C$_3$H$^+$, which are strongly linked to H$_2$CCCH$^+$, are shown in Fig.~\ref{Fig:model}.

As can be seen in Fig.~\ref{Fig:model}, the H$_2$CCC and H$_2$CCCH$^+$ abundances observed are relatively well reproduced by the model for a relatively early molecular cloud age around 2$\times$10$^5$ years with, however, a smaller H$_2$CCC/H$_2$CCCH$^+$ ratio than the one observed. Looking in more detail at the chemistry of H$_2$CCCH$^+$ and H$_2$CCC (see Fig.~3 of \citealt{RN8363}), it appears that H$_2$CCC is a product of H$_2$CCCH$^+$ (and also of c-C$_3$H$_3^+$) but that the flow of protonation of H$_2$CCC toward H$_2$CCCH$^+$ is a very minor pathway for the formation of H$_2$CCCH$^+$ which is almost essentially produced by the reaction C$_3$H$^+$ + H$_2$. This inverted link between H$_2$CCC and H$_2$CCCH$^+$ explains the unusually high MH$^+$/M ratio, compared to those cases in which the protonated form comes from the protonation of the neutral form \citep{Agundez2022}. 

Considering the link between C$_3$H$^+$, H$_2$CCCH$^+$ and C$_3$, it is interesting to see if the observations of H$_2$CCCH$^+$ (this work) combined to the observation of C$_3$H$^+$ \citep{RN11344} allow us to estimate the abundance of C$_3$, which in our dense cloud  model is the second carbon reservoir, accounting for up to 15\% of carbon. If C$_3$ does not react with atomic oxygen, as calculated by \cite{RN2549}, the protonation reactions of C$_3$ producing C$_3$H$^+$ are by far the main reactions of destruction of C$_3$ and of production of C$_3$H$^+$. As these protonation reactions control the destruction of C$_3$, the uncertainties on the rates affect the abundance of C$_3$, but does not change the flux of these reactions. The underestimation of C$_3$H$^+$ in the model is therefore not related to these protonation rates, 
but to the rate of the reaction C$_3$H$^+$ + H$_2$, which controls the destruction of C$_3$H$^+$ \citep{Maluendes1993,sav05}.
A decrease in the rate of the reaction C$_3$H$^+$ + H$_2$ at 10~K by a factor of 3 allows us to reproduce the abundance of C$_3$H$^+$ observed by \cite{RN11344}, as shown in Fig. \ref{Fig:model}.  This change in the rate coefficient does not affect the flux of the C$_3$H$^+$ + H$_2$ reaction and therefore does not affect the abundance of H$_2$CCCH$^+$, as long as the branching ratios to H$_2$CCCH$^+$ and c-C$_3$H$_3^+$ are not varied. Indeed, since \HC is mainly destroyed by the reaction with electrons, its abundance is controlled by the rate of this reaction, which is known only over the temperature range 172-489 K \citep{RN3797} with a temperature dependency inconsistent with theory. An increase in this rate at 10 K by a factor of 5, which is not impossible given the uncertainties, allows us to reproduce the observation for H$_2$CCCH$^+$ with a ratio between H$_2$CCC/H$_2$CCCH$^+$ very close to the observed one. Considering the uncertainties of the different chemical reactions linking C$_3$ to C$_3$H$^+$ and H$_2$CCCH$^+$, the observations of C$_3$H$^+$ \citep{RN11344} and H$_2$CCCH$^+$ (this work) validate the chemical scheme controlling the formation of cyclic and linear C$_3$H and C$_3$H$_2$ and the large gas-phase abundance of C$_3$ in TMC-1, around 10$^{-6}$ relative to H$_2$.

It would also be very interesting to know the abundance of the more stable cyclic isomer c-C$_3$H$_3^+$, 
which in the chemical model is predicted to be slightly more abundant (a factor of three) than \HC. 
This species has no dipole moment, but its deuterated version, c-C$_3$H$_2$D$^+$, 
has a low dipole moment of 0.225~D \citep{hua11} and its rotational spectrum has been recently measured 
in the 90-230 GHz frequency range in Cologne \citep{gupta23}. 
We searched for c-C$_3$H$_2$D$^+$ in the QUIJOTE line survey but at the current level of sensitivity, 
this species is not detected and we derive an upper limit to its column density of 4$\times$10$^{12}$ cm$^{-3}$. 
Assuming that the c-C$_3$H$_3^+$/c-C$_3$H$_2$D$^+$ ratio is 10, 
as found for the analog case of CH$_3$CCH with three equivalent H nuclei \citep{Cabezas2021}, 
the column density of c-C$_3$H$_3^+$ is $<$4$\times$10$^{13}$ cm$^{-2}$, which is not very meaningful.




\begin{acknowledgements}
      The work has been supported by an ERC advanced grant (MissIons:
101020583), the Deutsche Forschungsgemeinschaft (DFG) via SFB 956 (project ID
184018867), sub-project B2, and the Ger\"atezentrum ”Cologne Center for Terahertz
Spectroscopy” (DFG SCHL 341/15-1). W.G.D.P.S. thanks the Alexander von Humboldt
foundation for funding through a postdoctoral fellowship. 
We also acknowledge the support from the MICINN projects PID2020-113084GB-I00 and PID2019-106110GB-I00, the CSIC project ILINK+ LINKA20353, the ERC grant ERC-2013-Syg610256-NANOCOSMOS, and the Regional Computing Center of the Universität zu Köln (RRZK) for providing computing time on the DFG-funded high performance computing system CHEOPS.
\end{acknowledgements}

\bibliography{LIRTRAP.bib,sthorwirth_bibdesk}
\bibliographystyle{aa}


\begin{appendix}

\section{Structural calculations, internal coordinates}
Bond lengths are given in Å, angles in degrees.

\subsection{\ce{H2CCCH+}}
\begin{verbatim}
H2C3H+, CCSD(T)/cc-pwCVQZ
H
C 1 r1
X 2 rd 1 a90
C 2 r2 3 a90 1 d180
X 4 rd 2 a90 3 d0
C 4 r3 5 a90 2 d180
H 6 r4 4 a1 5 d0
H 6 r4 4 a1 5 d180
r1   =        1.073118130747218
rd   =        1.000000818629780
a90  =       90.000000000000000
r2   =        1.228490857165352
d180 =      180.000000000000000
d0   =        0.000000000000000
r3   =        1.346565030012042
r4   =        1.086025065701484
a1   =      120.382476682815607
\end{verbatim}

\begin{verbatim}
H2C3H+, CCSD(T)/ANO1
H
C 1 r1
X 2 rd 1 a90
C 2 r2 3 a90 1 d180
X 4 rd 2 a90 3 d0
C 4 r3 5 a90 2 d180
H 6 r4 4 a1 5 d0
H 6 r4 4 a1 5 d180
r1   =        1.074705266639338
rd   =        1.000000204657382
a90  =       90.000000000000000
r2   =        1.234261558515133
d180 =      180.000000000000000
d0   =        0.000000000000000
r3   =        1.351397370372584
r4   =        1.088292080166718
a1   =      120.373581553220575    
\end{verbatim}

\subsection{\ce{H2CCC}}

\begin{verbatim}
H2CCC, CCSD(T)/cc-pwCVQZ
C
C 1 r1
X 2 rd 1 a90
C 2 r2 3 a90 1 d180
H 4 r3 2 a1 3 d0
H 4 r3 2 a1 5 d180
r1   =        1.287645153115967
rd   =        1.000000000000000
a90  =       90.000000000000000
r2   =        1.327897213723068
d180 =      180.000000000000000
r3   =        1.083584615827809
a1   =      121.271597504509145
d0   =        0.000000000000000    
\end{verbatim}

\begin{verbatim}
H2CCC, CCSD(T)/ANO1
C
C 1 r1
X 2 rd 1 a90
C 2 r2 3 a90 1 d180
H 4 r3 2 a1 3 d0
H 4 r3 2 a1 5 d180
r1   =        1.294632885150227
rd   =        1.000000204657382
a90  =       90.000000000000000
r2   =        1.333030062830918
d180 =      180.000000000000000
r3   =        1.085894615830395
a1   =      121.302559724582778
d0   =        0.000000000000000    
\end{verbatim}


\begin{table*}[h]
\small
\centering
\caption{\label{tab:a1} Calculated and experimental spectroscopic parameters of  H$_2$CCC and \HC (in MHz). Equilibrium rotational constants calculated at the ae-CCSD(T)/cc-pwCVQZ level of theory, zero-point vibrational corrections $\Delta A_0$, $\Delta B_0$, $\Delta C_0$ and centrifugal distortion constants calculated at the
fc-CCSD(T)/ANO1 level. Best theoretical estimates (BE) for the rotational and centrifugal distortion constants $X$ of \HC are estimated 
as $X^{scaled}_{H_2CCCH^+}=\frac{X^{exp}_{H_2CCC}}
{X^{calc}_{H_2CCC}}\times X^{calc}_{H_2CCCH^+}$, i.e., using isoelectronic H$_2$CCC as a calibrator.
}
\begin{tabular}{@{\extracolsep{0pt}}lr@{}lr@{}l|r@{}lr@{}lr@{}l}
\hline
                     & \multicolumn{4}{c}{H$_2$CCC} & \multicolumn{4}{c}{\HC } \\ \cline{2-5} \cline{6-11}
                     &  \multicolumn{2}{c}{Experiment}               &     \multicolumn{2}{c}{Calculated} &     \multicolumn{2}{c}{Calculated}  &     \multicolumn{2}{c}{Scaled} &   \multicolumn{2}{c}{Experiment}                \\   
Parameter            &  \multicolumn{2}{c}{\cite{vrtilek_ApJL_364_L53_1990}} &  \multicolumn{2}{c}{This study}   & \multicolumn{2}{c}{This study} & \multicolumn{2}{c}{This study, BE} & \multicolumn{2}{c}{This study}  \\ 
\hline
$A_e$            & \multicolumn{2}{c}{$\cdots$}  & 292302 & .539 &  285650 & .316  & \multicolumn{2}{c}{$\cdots$} & \multicolumn{2}{c}{$\cdots$}  \\
$B_e$            & \multicolumn{2}{c}{$\cdots$}  &  10578 & .491 &    9675 & .219  & \multicolumn{2}{c}{$\cdots$} & \multicolumn{2}{c}{$\cdots$}  \\
$C_e$            & \multicolumn{2}{c}{$\cdots$}  &  10209 & .024 &    9358 & .247  & \multicolumn{2}{c}{$\cdots$} & \multicolumn{2}{c}{$\cdots$}  \\
$\Delta A_0$     & \multicolumn{2}{c}{$\cdots$}  &   2357 & .134 &    2331 & .884  & \multicolumn{2}{c}{$\cdots$} & \multicolumn{2}{c}{$\cdots$}  \\ 
$\Delta B_0$     & \multicolumn{2}{c}{$\cdots$}  &   $-$3 & .807 &       5 & .043  & \multicolumn{2}{c}{$\cdots$} & \multicolumn{2}{c}{$\cdots$}  \\
$\Delta C_0$     & \multicolumn{2}{c}{$\cdots$}  &     10 & .869 &      20 & .431  & \multicolumn{2}{c}{$\cdots$} & \multicolumn{2}{c}{$\cdots$}  \\
$A_0$            & 288783 & .(34)                  & 289945 & .405 &  283318 & .432  & 282182 & .595                &   281856 & .(247)    \\ 
$B_0$            &  10588 & .639(2)              &  10582 & .298 &    9670 & .175  &   9675 & .970                &     9675 & .841(1)  \\
$C_0$            &  10203 & .966(2)              &  10198 & .155 &    9337 & .816  &   9343 & .136                &     9342 & .877(1)  \\
$D_J\times 10^3$ &      4 & .248(2)                 &      3 & .722 &       2 & .684  &      3 & .063                &      2 & .942(4)    \\ 
$D_{JK}$         &      0 & .5164(5)                &      0 & .571 &       0 & .479  &      0 & .4328                &      0 & .4388(9)   \\ 
$D_K$            &     23 & .535$^a$             &     21 & .934 &      20 & .678  &    [20 & .678]               &     20 & .678$^a$   \\ 
$d_1\times 10^3$ &   $-$0 & .153(2)                 &   $-$0 & .143 &    $-$0 & .096  &   $-$0 & .103                &   $-$0 & .121(3)    \\   
$d_2\times 10^3$ &   $-$0 & .070(1)                 &   $-$0. & 050 &    $-$0 & .037  &   $-$0 & .051                &   $-$0 & .051$^a$   \\ 
%

\hline
\end{tabular}
  \tablefoot{\\
\tablefoottext{a}{Kept fixed in the analysis.}\\
 }
\end{table*}

\section{Line parameters of H$_2$CCCH$^+$ and H$_2$CCC} 

\begin{figure*}
   \centering
\includegraphics[width=0.9\textwidth]{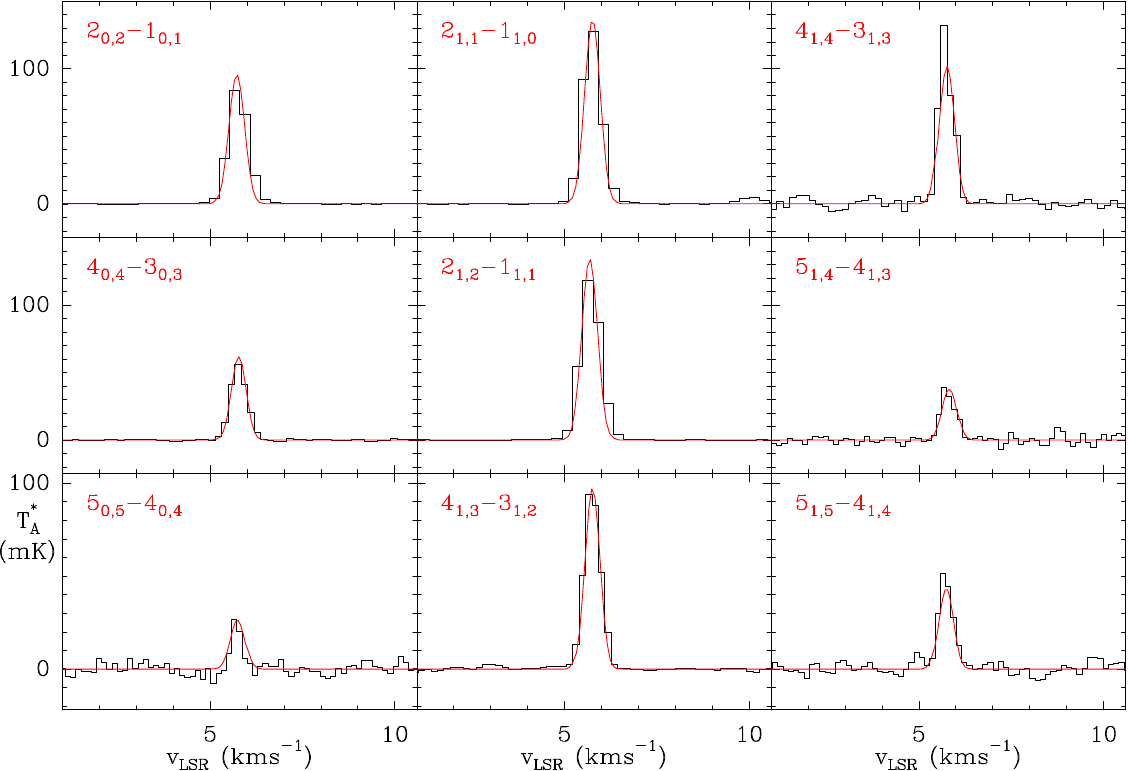}
   \caption{Observed lines of H$_2$CCC in the 31–115~GHz
domain towards TMC-1. The abscissa corresponds to the velocity of the cloud with
respect to the Local Standard of Rest (v$_{LSR}$).
 The ordinate is the antenna temperature corrected for atmospheric and telescope losses in mK. The spectral resolution is 38~kHz below 50~GHz and 48~kHz above. Derived line parameters are given in Table \ref{Table:Line_Parameters_H2C3}. The red lines show the computed synthetic spectra
for the lines of H$_2$CCC (see text). The lines in the left column correspond to para lines of H$_2$CCC, while those
of the central and right columns correspond to transitions of the ortho species. The adopted frequencies are those of the
CDMS catalogue \citep{Muller2005} and are given in Table \ref{Table:Line_Parameters_H2C3}.}
\label{FigH2C3}%
    \end{figure*}

The line parameters of H$_2$CCCH$^+$ and H$_2$CCC have been obtained by fitting a Gaussian
line profile to the observed data. The results are given in Tables \ref{Table:Line_Parameters}
and \ref{Table:Line_Parameters_H2C3}, respectively. The observed lines of H$_2$CCCH$^+$ are shown
in Fig. \ref{FigGam} and those of H$_2$CCC in Fig. \ref{FigH2C3}.

\onecolumn

\begin{table*}
\centering 
\caption{Line parameters of the observed transitions of H$_2$CCCH$^+$ in TMC-1.}
    \label{Table:Line_Parameters}
    \begin{tabular}{lcccccc}
    \hline\hline
Transition & $\nu_{obs}$$^a$ & $\int$ $T_A^*$ dv $^b$  & $\Delta$v$^c$ & T$_A^*$$^d$ & $\sigma$$^e$ & Notes\\
           & (MHz)          & (mK $\times$ kms$^{-1}$)& (kms$^{-1}$) & (mK) & (mK)\\
    \hline
$2_{1,2}-1_{1,1}$ & 37702.627$\pm$0.010 & 0.38$\pm$0.13 & 0.63$\pm$0.20 & 0.56 & 0.13 &  \\
$2_{0,2}-1_{0,1}$ & 38037.044$\pm$0.010 & 0.36$\pm$0.09 & 0.78$\pm$0.17 & 0.43 & 0.13 &  \\
$2_{1,1}-1_{1,0}$ & 38368.581$\pm$0.010 & 0.57$\pm$0.07 & 0.77$\pm$0.11 & 0.70 & 0.13 & A\\
$4_{0,4}-3_{0,3}$ & 76071.062$\pm$0.020 &               &               &      & 3.40 & B\\
$4_{1,3}-3_{1,2}$ & 76735.888$\pm$0.020 & 2.28$\pm$0.07 & 0.40$\pm$0.10 & 5.50 & 1.80 & CD\\
$5_{1,5}-4_{1,4}$ & 94254.055$\pm$0.004 &               &               &      & 2.04 & B\\
$5_{0,5}-4_{0,4}$ & 95085.988$\pm$0.030 & 3.27$\pm$0.08 & 0.85$\pm$0.23 & 3.60 & 1.40 & BD\\
$5_{1,4}-4_{1,3}$ & 95918.765$\pm$0.020 & 1.04$\pm$0.20 & 0.40$\pm$0.10 & 3.30 & 0.67 &  \\
$6_{1,6}-5_{1,5}$ &113103.272$\pm$0.004 &               &               &      & 4.40 & B\\
$6_{0,6}-5_{0,5}$ &114099.086$\pm$0.006 &               &               &      & 4.90 & B\\
    \hline
    \hline
    \end{tabular}
\tablefoot{\\
\tablefoottext{a}{Observed frequency assuming a v$_{LSR}$ of 5.83 km\,s$^{-1}$.}
\tablefoottext{b}{Integrated line intensity in mK\ $\times$ km\,s$^{-1}$.}
\tablefoottext{c}{Line width in km\,s$^{-1}$.}
\tablefoottext{d}{Antenna temperature in mK.}
\tablefoottext{e}{Root mean square noise of the data.}
\tablefoottext{A}{Blended with a weak feature at 38638.7 MHz.}
\tablefoottext{B}{For undetected lines their rest frequencies correspond to the observed frequency in
the laboratory (Table \ref{rotlines}), or to the predicted one from the molecular constants of
Table \ref{tab:fit}.}
\tablefoottext{C}{The feature appears too strong.}
\tablefoottext{D}{Marginal detection.}
}
\end{table*}

\begin{table*}
\centering 
\caption{Line parameters of the observed transitions of H$_2$CCC in TMC-1.}

    \label{Table:Line_Parameters_H2C3}
    \begin{tabular}{lccccccc}
    \hline\hline
Transition & $\nu_{obs}$$^a$ & $\int$ $T_A^*$ dv $^b$  & v$_{LSR}$$^c$ & $\Delta$v$^d$ & T$_A^*$$^e$ & $\sigma$$^f$ \\
           & (MHz)          & (mK $\times$ kms$^{-1}$)& (kms$^{-1}$) &  (kms$^{-1}$) & (mK) & (mK)\\
    \hline
$2_{1,2}-1_{1,1}$& 41198.335$\pm$0.002& 82.7$\pm$0.1 & 5.70$\pm$0.01 &  0.65$\pm$0.01 &  120.4 & 0.1\\
$2_{0,2}-1_{0,1}$& 41584.675$\pm$0.001& 57.8$\pm$0.1 & 5.76$\pm$0.01 &  0.63$\pm$0.01 &   86.9 & 0.2\\
$2_{1,1}-1_{1,0}$& 41967.671$\pm$0.002& 84.6$\pm$0.1 & 5.75$\pm$0.01 &  0.61$\pm$0.01 &  130.8 & 0.2\\
$4_{1,4}-3_{1,3}$& 82395.089$\pm$0.003& 61.9$\pm$1.3 & 5.71$\pm$0.01 &  0.47$\pm$0.01 &  124.8 & 3.0\\
$4_{0,4}-3_{0,3}$& 83165.345$\pm$0.003& 31.9$\pm$0.4 & 5.77$\pm$0.02 &  0.54$\pm$0.01 &   55.5 & 0.9\\
$4_{1,3}-3_{1,2}$& 83933.699$\pm$0.003& 56.6$\pm$0.4 & 5.75$\pm$0.02 &  0.54$\pm$0.01 &   97.9 & 0.9\\
$5_{1,5}-4_{1,4}$&102992.379$\pm$0.004& 25.4$\pm$1.2 & 5.73$\pm$0.02 &  0.47$\pm$0.03 &   51.1 & 3.1\\
$5_{0,5}-4_{0,4}$&103952.926$\pm$0.003& 86.7$\pm$1.1 & 5.71$\pm$0.03 &  0.29$\pm$0.05 &   28.0 & 3.6\\
$5_{1,4}-4_{1,3}$&104915.583$\pm$0.003& 19.0$\pm$1.1 & 5.77$\pm$0.03 &  0.48$\pm$0.03 &   37.4 & 2.9\\
\hline
\end{tabular}
\tablefoot{\\
\tablefoottext{a}{Adopted rest frequencies from the CDMS entry for H$_2$CCC.}
\tablefoottext{b}{Integrated line intensity in mK $\times$ km\,s$^{-1}$.}
\tablefoottext{c}{Velocity of the line in km\,s$^{-1}$.}
\tablefoottext{d}{Line width in km\,s$^{-1}$.}
\tablefoottext{e}{Antenna temperature in mK.}
\tablefoottext{f}{Root mean square noise of the data.}
}
\end{table*}

\end{appendix}

\end{document}